\newcommand{\idty}{{\leavevmode{\rm 1\mkern -5.4mu I}}}
\newcommand{\Ir}{Z\!\!\!Z}
\newcommand{\Ibb}[1]{ {\rm I\ifmmode\mkern
            -3.6mu\else\kern -.2em\fi#1}}
\newcommand{\ibb}[1]{\leavevmode\hbox{\kern.3em\vrule
     height 1.2ex depth -.3ex width .2pt\kern-.3em\rm#1}}
\newcommand{\Cx}{{\ibb C}}
\newcommand{\Rl}{{\Ibb R}}
\documentstyle[12pt]{article}
\begin{document}
\renewcommand{\theequation} {\arabic{section}.\arabic{equation}}
\begin{tabbing}
\hspace*{12cm}\= GOET-TP 33/93 \\
              \> revised  \\
              \> November 1993
\end{tabbing}
\vspace*{2cm}
\centerline{\huge \bf Differential Calculus and Gauge Theory}
\vskip.5cm
\centerline{\huge \bf on Finite Sets ${}^\ast$}
\vskip1.5cm

\begin{center}
      {\bf Aristophanes Dimakis}
      \vskip.1cm
      Department of Mathematics, University of Crete \\
      GR-71409 Iraklion, Greece
      \vskip.2cm
       and
      \vskip.2cm
      {\bf Folkert M\"uller-Hoissen}
      \vskip.1cm
      Institut f\"ur Theoretische Physik  \\
      Bunsenstr. 9, D-37073 G\"ottingen, Germany
\end{center}

\vskip1.5cm

\begin{abstract}
\noindent
We develop differential calculus and gauge theory on a finite set $G$.
An elegant formulation is obtained when $G$ is supplied with a group
structure and in particular for a cyclic group.
Connes' two-point
model (which is an essential ingredient of his reformulation of the
standard model of elementary particle physics) is recovered in our
approach. Reductions of the universal differential calculus
to `lower-dimensional' differential calculi are considered. The
`complete reduction' leads to a differential calculus on a
periodic lattice.
\end{abstract}
\vskip1cm
\centerline{to appear in {\em Journal of Physics A: Math. Gen.}}

\vskip2cm
${}^\ast$ Work supported by the Heraeus-Foundation.

\newpage

\section{Introduction}
\setcounter{equation}{0}
Again and again over the years, arguments have been given to assign a
more fundamental role to discrete spaces rather than to the continuum
and attempts were made to develop corresponding physical theories
(see \cite{Amba30} for some early examples). Such an idea is being
pursued by David Finkelstein since 1968 \cite{Fink69} culminating
in a forthcoming book. Classical and
quantum field theory on discrete spaces has been considered, in
particular, in \cite{Yama84}. The finiteness of the entropy of a black
hole (and the corresponding finiteness of the number of bits of
information that can be stored there) led `t Hooft to speculate about
a discrete (cellular automaton) structure of space-time at the Planck
scale \cite{tHoo90}. Further interesting ideas about discreteness of
space and time can be found in \cite{Feyn82,BLMS87}, for example.
\vskip.2cm

More recently, concepts of differential geometry were extended
to discrete spaces (and even noncommutative algebras).
In the framework of noncommutative geometry, finite spaces have
been considered to build models of elementary particle
physics \cite{Conn} (see also \cite{noncom-geom}).
The present work provides a general approach to the differential
geometry of such spaces. It has been inspired by recent papers of
Sitarz \cite{Sita92} who treated the case of discrete groups
(see also \cite{Conn86}).
\vskip.2cm

We take the point of view that some form of differential calculus
is the very basic structure to formulate physical models and, in
particular, dynamics of fields on some space. Our belief in the
physical relevance of this mathematical structure is partly
based on the observation made in \cite{DMH92,DMHS93} that lattice
theories (in particular their Lagrangian and action) are obtained
from continuum theories in a universal way simply by a certain
deformation of the ordinary calculus of differential forms. In
this case functions and differentials satisfy non-trivial
commutation relations depending on the lattice spacings (for
vanishing lattice spacings they commute and one recovers the
ordinary differential calculus). Another deformation of the
ordinary differential calculus was shown to be related with
stochastic calculus \cite{DMH93} and `proper time' formulations
\cite{Fanc93} of physical theories \cite{DMH92b}.
\vskip.2cm

In section 2 we introduce the universal differential calculus
({\em universal differential envelope} \cite{Conn86,Coqu89}) on
an arbitrary finite set of $N$ elements.
Section 3 shows how to formulate gauge theory on a finite set.
Of special interest is the case when the set is supplied with
a group structure. This is the subject of section 4.
In section 5, we consider the group structure $\Ir_N$ in detail.
\vskip.2cm

In the gauge theory formalism based on the universal differential
calculus the connection (gauge potential) on a finite set plays
the role of (a set of) Higgs fields \cite{Conn}.
This observation stimulated the use of noncommutative geometry for
model building in particle physics \cite{Conn,noncom-geom}.
We briefly discuss it for the two-point space in section 5.2.
\vskip.2cm

The universal differential calculus on a finite set of order $N$
associates with it $N-1$ linearly independent differentials.
On the other hand, we know that there are `smaller' differential
calculi. In particular, it is possible to have the $N$ points
`on a closed line', i.e. -- embedded as a lattice -- in one
dimension. This configuration is described by a single coordinate
$y$ which satisfies the commutation relation
\begin{eqnarray*}
                     y \, dy = q \, dy \, y
\end{eqnarray*}
with its differential $dy$ where $q$ is an $N$-th primitive root
of unity (see \cite{DMH92}).
There is a natural way from the universal differential calculus to
this `reduced' differential calculus. Related with the fact that one
always has the group structure $\Ir_N$ on a set of $N$ elements, there
is a function $y$ such that $y^N = 1$. Expressing the universal
differential calculus in terms of the functions $y^n$, $n=0,\ldots,N-1$,
we can consistently add the above relation so that the
($N-1$)-dimensional universal calculus is
reduced to a one-dimensional differential calculus. Details
are presented in section 6. The reduced differential calculus (and its
higher-dimensional generalization) gives a convenient universal
framework to formulate and describe physical models on a (closed)
lattice \cite{DMH92,DMHS93}.
\vskip.2cm

Between the universal differential calculus (which assigns an
($N-1$)-dimensional polyhedron to a set of $N$ points) and the
one-dimensional (periodic lattice) calculus there are other
differential calculi. It is our concern what kind of
geometric\footnote{The notion `geometric' here refers to a connection
structure on the set of points in the sense of graphs. We do not
consider a metric (or distance function) on the point set in this work.}
structures can be associated with them. Section 6 explores some
of the possibilities.
\vskip.2cm

A reduction of the universal differential calculus induces a
corresponding reduction of structures built on it, like gauge theory.
In this way one can approach field theory on finite sets.
\vskip.2cm
\noindent
Section 7 contains some conclusions.

\section{Differential calculus on a finite set}
\setcounter{equation}{0}
Let $G$ be a finite set with $N$ elements and $\cal A$ the algebra
of $\Cx$-valued functions on $G$ with the usual pointwise
multiplication of functions,
\begin{eqnarray}
     (f f')(g) = f(g) \, f'(g)   \qquad \forall f,f'\in {\cal A},
                                        \; g \in G  \; .
\end{eqnarray}
${\cal A}$ is a commutative, associative and unital complex algebra.
With each $g \in G$ we associate a function $x_g \in {\cal A}$ such
that
\begin{eqnarray}         \label{x_g}
    x_g(g') = \delta_{g,g'}     \qquad \forall g' \in G  \; .
\end{eqnarray}
The functions $x_g$ satisfy the identities
\begin{eqnarray}
 \sum_{g \in G} \, x_g &=& \idty  \qquad
        (\mbox{the unit in $\cal A$})         \label{sum=1} \\
         x_g \, x_{g'} &=& \delta_{g g'} \, x_g
                        \label{xx-x}
\end{eqnarray}
which will be frequently used in the following. As a consequence of
these two identities, every function $f \in {\cal A}$ has an
expansion
\begin{eqnarray}             \label{f-expansion}
           f = \sum_g \, f(g) \, x_g
\end{eqnarray}
which shows that the functions $x_g$ span $\cal A$ linearly over
$\Cx$. Furthermore,
\begin{eqnarray}
   x_g \, f &=& x_g \, f(g)   \qquad \forall f \in {\cal A}  \; .
                                                   \label{x_g-f}
\end{eqnarray}
The complex conjugate of $f \in {\cal A}$ will be denoted
as $f^\ast$.
\vskip.3cm

We extend ${\cal A} =: \Omega^0$ to a differential algebra via the
action of an {\em exterior derivative} operator $d$. It maps elements of
$\cal A$ into (formal) differentials which span the space $\Omega^1$ of
1-forms as an $\cal A$-bimodule, and furthermore $r$-forms into
$(r+1)$-forms, i.e. $d \, : \, \Omega^r \rightarrow \Omega^{r+1}$.
We require
\begin{eqnarray}
            d \idty &=& 0   \label{didty} \\
                d^2 &=& 0    \\
          d(\omega \omega') &=& d\omega \; \omega' + (-1)^r \,
                                \omega \, d \omega'
                                \label{Leibniz}
\end{eqnarray}
where $\omega$ and $\omega'$ are $r$- and $r'$-forms, respectively.
$\Omega$ denotes the space $\bigoplus_{r=0}^\infty \Omega^r$ of all
forms.
\vskip.3cm

Now (\ref{sum=1}) implies
\begin{eqnarray}              \label{sum-dx}
     \sum_g \, dx_g = 0   \; ,
\end{eqnarray}
(\ref{f-expansion}) leads to
\begin{eqnarray}              \label{df-expansion}
    df = \sum_g \, f(g) \, dx_g
\end{eqnarray}
since $f(g)$ are constants, and (\ref{x_g-f}) yields
\begin{eqnarray}              \label{f-dx}
          f \, dx_g = f(g) \, dx_g - df \, x_g    \; .
\end{eqnarray}
Acting with $d$ on these relations does not lead to further relations.
(\ref{f-dx}) has the form of commutation relations between differentials
and elements of $\cal A$. It should be noticed that these relations
are simply a consequence of the Leibniz rule (\ref{Leibniz}).
\vskip.3cm

We will now choose an element $e \in G$ once and for all.\footnote{In
the case of a group a natural choice will be the unit element. In
general, however, there will not be a distinguished choice of $e$.}
(\ref{sum-dx}) shows that the differentials $dx_g$ are
linearly dependent. The differentials $dx_g$ with $g \in G \setminus
\lbrace e \rbrace =: G'$ are linearly independent, however.
Instead of $\sum_{g \in G'}$ we will write $\sum_g'$ in the
following. Then
\begin{eqnarray}      \label{df=sum-difference}
    df = {\sum_g}' \, dx_g \, \lbrack f(g) - f(e) \rbrack
\end{eqnarray}
which shows that $df=0$ iff $f$ takes the same value at each
element of $G$.
\vskip.3cm

The $x_g$ are real functions, i.e,
\begin{eqnarray}           \label{x-real}
    (x_g)^\ast = x_g \; .
\end{eqnarray}
The complex conjugation on $\cal A$ can be extended to an involution
of the differential algebra by \cite{Conn}
\begin{eqnarray}            \label{involution}
  (f_1 \, df_2 \cdots df_k)^\ast = d (f_k^\ast) \cdots
     d (f_2^\ast) \, f_1^\ast
\end{eqnarray}
where $f_\ell \in {\cal A}$.
\vskip.3cm

If we put an ordering on the elements of $G$, i.e. $g_0, \ldots,
g_{N-1}$, then there is another natural choice of an involution on
$\cal A$. It is determined by
\begin{eqnarray}                  \label{x-star}
    (x_i)^\star = x_{N-i}
\end{eqnarray}
where $x_i := x_{g_i}$. Again, it extends to $\Omega({\cal A})$ via
\begin{eqnarray}            \label{star-involution}
  (f_1 \, df_2 \cdots df_k)^\star = d (f_k^\star) \cdots
     d (f_2^\star) \, f_1^\star    \; .
\end{eqnarray}
On $\Cx$, the involution should coincide with complex conjugation.
We will find the $\star$-involution of importance when we
consider reductions of the differential calculus in section 6.
\vskip.3cm

Let us introduce the 1-forms
\begin{eqnarray}
            \theta_{ij} = dx_i \, x_j
\end{eqnarray}
for $i \neq j$. It follows that
\begin{eqnarray}
 \theta_{jk} \, x_i = \theta_{jk} \, \delta_{ki} \quad , \quad
 x_i \, \theta_{jk} = \delta_{ij} \, \theta_{jk}
    \qquad (j \neq k)
\end{eqnarray}
so that the $\theta_{jk}$ are common eigen-1-forms for all $x_i$.
As a consequence, if we impose the condition $\theta_{jk} = 0$ for
fixed $j$ and $k$ on the differential calculus, then we do not
obtain further relations by multiplication with elements of $\cal A$.
We shall return to this observation in section 6.
The $\theta_{ij}$ are a basis of the space of 1-forms as a complex
vector space. In particular, we have
\begin{eqnarray}
  df = \sum_{i \neq j} \theta_{ij} \, \lbrack f(g_i) - f(g_j) \rbrack
     = \lbrack - \sum_{i \neq j} \theta_{ij} \, , \, f \rbrack
\end{eqnarray}
for $f \in \cal A$.
The two involutions introduced above act on $\theta_{ij}$ in the
following way,
\begin{eqnarray}
      \theta_{ij}^\ast &=& - \theta_{ji}            \\
      \theta_{ij}^\star &=& - \theta_{{N-j},{N-i}}  \; .
\end{eqnarray}
\vskip.3cm

The simple relations which the $x_i$ and the $\theta_{jk}$ satisfy
suggest the following construction of finite-dimensional matrix
representations of the differential calculus. Let $E_{ij}$ denote
the $N\times N$-matrix with components $(E_{ij})_{k \ell} =
\delta_{ik} \delta_{j \ell}$. The matrices
\begin{eqnarray}
 \rho(x_i) = \left(\begin{array}{cc}
   E_{ii} & 0       \\
   0      & E_{ii} \end{array}\right)
      \quad , \quad
 \rho(\theta_{ij}) = \left(\begin{array}{cc}
   0 & C_{ij}  \, E_{ij}  \\  C'_{ij}  \, E_{ij} & 0
                     \end{array}\right)     \label{reps}
\end{eqnarray}
with (non-zero) constants $C_{ij}, C'_{ij}$ yield a representation of
the differential algebra. However, only functions and 1-forms are
{\em faithfully} represented. If $C^\ast_{ij} = - C'_{ji}$,
our first involution acts by Hermitian conjugation on the
matrices. To represent the second involution, an additional
`reflection' of the indices has to be performed. The `doubling' of the
matrices in (\ref{reps}) is necessary for $N > 2$ to account for the
$\Ir_2$-grading of the differential algebra. The case $N=2$ is special
in this respect (see also appendices A and B).
More general representations are given by
\begin{eqnarray}
 \rho (x_i) = \left(\begin{array}{cc}
   {\bf 1}_M \otimes E_{ii} & 0       \\
                     0      & {\bf 1}_M \otimes E_{ii}
                     \end{array}\right)
      \quad , \quad
 \rho (\theta_{ij}) = \left(\begin{array}{cc}
   0 & C_{ij} \otimes E_{ij}  \\  C'_{ij} \otimes E_{ij} & 0
                       \end{array}\right)
\end{eqnarray}
where $C_{ij}$ and $C'_{ij}$ are now $M\times M$-matrices and
${\bf 1}_M$ denotes the $M\times M$ unit matrix.

\section{Gauge fields on a finite set}
\setcounter{equation}{0}
Using the differential calculus introduced in the previous section,
gauge theory can now be formulated on a finite set $G$.
Section 3.1 deals with the concept of a connection and
its field strength (curvature). In section 3.2 we
consider the covariant derivative of a field which transforms
according to a representation of the gauge group. If a conjugation
is given on the space of fields such that it is compatible with
the connection, then the connection turns out to be anti-Hermitian
for a unitary gauge group, in analogy with the continuum case.

\subsection{Connection and field strength}
Let
\begin{eqnarray}          \label{A}
    A = \sum_g \, dx_g \, A_g
\end{eqnarray}
be a {\em connection 1-form} which transforms under a gauge
transformation according to the familiar rule
\begin{eqnarray}      \label{gauge-trafo}
    A' = U \, A \, U^{-1} - dU \, U^{-1}
\end{eqnarray}
where $U$ is a function on $G$ with values in a matrix group. Inserting
(\ref{A}) in (\ref{gauge-trafo}) and using (\ref{f-dx}) we find
\begin{eqnarray}             \label{dU...}
   dU \, ( 1 + \sum_g \, x_g \, A_g ) = \sum_g \, dx_g \,
   \lbrack U(g) \, A_g - A_g' \, U \rbrack   \; .
\end{eqnarray}
This equation is satisfied when\footnote{The last equation may admit
other solutions. This will not be further discussed in the
present work. Note that the $1$ appearing in the expression
(\ref{dU...}) and on the right hand sides of (\ref{sum-x-A}) and
(\ref{sum-gauge-inv}) has to be understood as $\idty$ times the unit
matrix of the gauge group.}
\begin{eqnarray}
    \sum_g \, x_g \, A_g &=& -1   \label{sum-x-A}  \\
    A_g' &=& U(g) \, A_g \, U^{-1}   \; .
\end{eqnarray}
One has to check that (\ref{sum-x-A}) is gauge-invariant. Indeed,
\begin{eqnarray}          \label{sum-gauge-inv}
  \sum_g \, x_g \, A_g' = \sum_g \, x_g \, U(g) \, A_g \, U^{-1}
                        = U \, \sum_g \, x_g \, A_g \, U^{-1}
                        = - U \, U^{-1} = -1  \; .
\end{eqnarray}
Because of (\ref{sum-dx}) the coefficients $A_g$ in (\ref{A})
are not uniquely determined by the left hand side of (\ref{A}).
The corresponding freedom is fixed, however, by the condition
(\ref{sum-x-A}).
\vskip.2cm

Multiplying (\ref{sum-x-A}) by $x_{g'}$ and using (\ref{x_g-f}) yields
\begin{eqnarray}
                x_g \, ( A_g(g) + 1 ) = 0
\end{eqnarray}
so that
\begin{eqnarray}       \label{A_g=-1}
                A_g(g) = -1    \qquad  \forall g \in G  \; .
\end{eqnarray}
    From (\ref{gauge-trafo}) one finds that
\begin{eqnarray}
 F = dA + A \, A = \sum_{g,g'} \, dx_g \, dx_{g'} \, A_g(g') \, A_{g'}
\end{eqnarray}
transforms according to $F \mapsto U \, F \, U^{-1}$.

\subsection{Covariant derivatives}
Using (\ref{gauge-trafo}) we have
\begin{eqnarray}
  U \, dx_g = dx_g \, U(g) - dU \, x_g
            = dx_g \, U(g) - U \, A \, x_g + A' \, x_g \, U(g)
\end{eqnarray}
which shows that
\begin{eqnarray}
                Dx_g := dx_g + A \, x_g
\end{eqnarray}
transforms covariantly,
\begin{eqnarray}
                D'x_g = U \, Dx_g \, U(g)^{-1}   \; .
\end{eqnarray}
Furthermore, it satisfies
\begin{eqnarray}       \label{sum-Dx-A}
                \sum_g \, Dx_g = A  \; .
\end{eqnarray}
Let us consider a field $\psi$ on $G$ as an element of a module
${\cal V} := {\cal A}^n$. If the gauge group acts on it
according to $ \psi \mapsto U \, \psi$, then
\begin{eqnarray}
               D \psi := d \psi + A \, \psi
\end{eqnarray}
has the same transformation property as a consequence of
(\ref{gauge-trafo}). Using (\ref{f-expansion}) one finds
\begin{eqnarray}
          D \psi = \sum_g \, Dx_g \, \psi(g)  \; .
\end{eqnarray}
     From (\ref{sum-Dx-A}) and (\ref{sum-x-A}) one obtains the identity
\begin{eqnarray}
          \sum_g \, Dx_g \, A_g = 0  \; .
\end{eqnarray}
Furthermore, using the Leibniz rule for $d$, (\ref{x_g-f}),
(\ref{A_g=-1}) and (\ref{sum=1}),
one can express $D \psi$ as a right-form,
\begin{eqnarray}
 D \psi = {\sum_g}' \, \stackrel{\leftarrow}{\nabla}_g \psi \; dx_g
\end{eqnarray}
with
\begin{eqnarray}
  \stackrel{\leftarrow}{\nabla}_g \psi = \sum_{g'} \, x_{g'} \,
  \lbrack A_{g'}(e) \, \psi(e) - A_{g'}(g) \, \psi(g) \rbrack \; .
\end{eqnarray}
\vskip.3cm

Let a conjugation be given which maps $\psi \in {\cal V}$ to
an element $\psi^\dagger$ of the dual module
${\cal V}^\ast$ such that $(\omega \psi)^\dagger = \psi^\dagger
\omega^\ast$ (or with $\omega^\ast$ replaced by $\omega^\star$).
If $U$ is unitary, i.e. $U^\dagger = U^{-1}$, then the assumption
\begin{eqnarray}
                (D \psi)^\dagger = D (\psi^\dagger)
\end{eqnarray}
implies
\begin{eqnarray}
  d \psi^\dagger + \psi^\dagger \, A^\dagger = (d \psi + A \,
  \psi)^\dagger = d \psi^\dagger - \psi^\dagger \, A
\end{eqnarray}
and therefore
\begin{eqnarray}        \label{A-dagger}
                    A^\dagger = - A     \; .
\end{eqnarray}

\section{Differential calculus and gauge theory on a finite group}
\setcounter{equation}{0}
Differential geometric structures on Lie groups like Maurer-Cartan
forms play an important role in the construction of physical models
and in particular in the formulation of gauge theories as structures
on principal fiber bundles. It is therefore of interest that these
structures can also be formulated on discrete groups (see also
\cite{Sita92}).
\vskip.2cm

Since on a finite set $G$ it is always possible to introduce a group
structure, such a structure can be used to rewrite the differential
calculus and gauge theory introduced in the previous sections in a
different and often simpler form. This will be shown in the
following.

\subsection{Differential calculus on a finite group}
Let us consider a group structure on a finite set $G$ with
group multiplication $(g,g') \mapsto g g'$.
For the element $e \in G$ we choose the unit element of the group.
Right and left actions on the algebra $\cal A$ of functions on $G$
are then given by
\begin{eqnarray}
  (R_g f)(g') := f(g' g) \quad , \quad
  (L_g f)(g') := f(g g')                 \; .
\end{eqnarray}
One finds that the 1-forms
\begin{eqnarray}      \label{theta}
   \theta_g := \sum_{g'} \, dx_{g'g} \, x_{g'}
             = \sum_{g'} \, dx_{g'} \, x_{g'g^{-1}}
\end{eqnarray}
are left-invariant, i.e.
\begin{eqnarray}
  L_{g''} \theta_g &:=& \sum_{g'} \, dx_{ {g''}^{-1} g'g}
                        \, x_{{g''}^{-1} g'} = \theta_g      \; .
\end{eqnarray}
The 1-forms $\theta_g$ are in this sense analogues of left-invariant
Maurer-Cartan forms on a Lie group. They satisfy the identity
\begin{eqnarray}            \label{sum-theta}
                  \sum_g \, \theta_g = 0
\end{eqnarray}
as a consequence of (\ref{sum-dx}). Furthermore,
\begin{eqnarray}                 \label{dx=}
   dx_g = \sum_{g'} \, \theta_{g'} \, x_{g {g'}^{-1}}   \; .
\end{eqnarray}
Using (\ref{theta}), (\ref{f-dx}), (\ref{x_g-f}) and (\ref{sum=1}),
one derives the relation
\begin{eqnarray}       \label{f-theta}
  f \, \theta_g = \theta_g \, R_g f - \delta_{g,e} \, df   \; .
\end{eqnarray}
As a consequence of (\ref{f-theta}),
\begin{eqnarray}
      df = \lbrack \theta_e \, , \, f \rbrack
\end{eqnarray}
which assigns a particular role to $\theta_e$.
Furthermore, we obtain
\begin{eqnarray}
      df &=& \sum_g \, \theta_g \, R_g f
          = {\sum_g}' \, \theta_g \, (R_g - 1) f
                    \label{df-theta}
\end{eqnarray}
using (\ref{f-expansion}), (\ref{dx=}), (\ref{x_g-f}), (\ref{sum=1})
and (\ref{sum-theta}).
Acting with $d$ on (\ref{theta}) leads to
\begin{eqnarray}
  d \theta_g &=& \sum_{g'} \, \theta_{g'} \, \theta_{g {g'}^{-1}}
                      \label{dtheta}
\end{eqnarray}
which resemble Maurer-Cartan equations.
The expression obtained by acting with $d$ on (\ref{df-theta}), using
(\ref{dtheta}) and (\ref{df-theta}), has to vanish as a consequence
of $d^2 =0$. It vanishes identically, however, so that we do not obtain
relations between 1-forms.
\vskip.3cm

Using (\ref{theta}), (\ref{x-real}), (\ref{involution}),
(\ref{Leibniz}), (\ref{x_g-f}) and (\ref{sum-dx}), one finds
\begin{eqnarray}            \label{theta-ast}
                (\theta_g)^\ast = - \theta_{g^{-1}}  \; .
\end{eqnarray}
In case of the $\star$-involution (cf section 2) one obtains
\begin{eqnarray}            \label{theta-star}
                (\theta_g)^\star = - \theta_g   \; .
\end{eqnarray}
\vskip.3cm
It is also possible, of course, to introduce {\em right}-invariant
Maurer-Cartan forms. All the above relations for $\theta_g$ have
corresponding counterparts.
\vskip.3cm
If the number $N$ of elements of $G$ is a prime number, then the
only possible group structure is $\Ir_N$ \cite{Hame62}. In the general
case, each group with $N$ elements must be a subgroup of the symmetric
(permutation) group $S_N$. For a finite set, the symmetric group
is of particular interest since it plays the role of the homeomorphism
(or diffeomorphism) group of topological spaces or manifolds.
The symmetric group and its representations are therefore expected
to be important in an approach towards a theory of gravity on a
finite discrete space-time.

\subsection{Gauge fields on a finite group}
Let us write the connection 1-form $A$ in terms of the
1-forms $\theta_g$,
\begin{eqnarray}       \label{A-theta}
     A = \sum_g \, \theta_g \, P_g   \; .
\end{eqnarray}
Then
\begin{eqnarray}
     A_{g'} = \sum_g \, x_{g' g^{-1}} \, P_g
\end{eqnarray}
and the condition (\ref{sum-x-A}) translates into the much simpler
equation
\begin{eqnarray}    \label{P_e=-1}
     P_e = -1
\end{eqnarray}
(where the $1$ stands for $\idty$ times the unit matrix of the gauge
group). Under a gauge transformation,
\begin{eqnarray}
     P_g' = (R_g U) \, P_g \, U^{-1}   \; .
\end{eqnarray}
Using (\ref{f-theta}) and (\ref{gauge-trafo}), the transformation
of the 1-forms
\begin{eqnarray}
    \tilde{\theta}_g := \theta_g + \delta_{g,e} \, A
\end{eqnarray}
is found to be given by
\begin{eqnarray}
   U \, \tilde{\theta}_g = \tilde{{\theta}'}_g \, R_g U \; .
\end{eqnarray}
They satisfy
\begin{eqnarray}
    \sum_g \tilde{\theta}_g = A  \; .
\end{eqnarray}
The field strength of $A$ is
\begin{eqnarray}
   F = \sum_{g,g'} \, \theta_g \, \theta_{g'} \, \lbrack
       P_{g g'} + (R_{g'}P_g) \, P_{g'} \rbrack
     = \sum_{g,g'} \, \tilde{\theta}_g \, \tilde{\theta}_{g'}
       \, \lbrack P_{g g'} + (R_{g'}P_g) \, P_{g'} \rbrack   \; .
\end{eqnarray}
For the covariant derivative of $\psi$ we obtain the expression
\begin{eqnarray}
  D \psi = \sum_{g} \, \tilde{\theta}_g \, \lbrack R_g \psi
           + P_g \psi \rbrack   \; .
\end{eqnarray}
With the $\ast$-involution, using (\ref{A-theta}) and (\ref{theta-ast}),
the condition (\ref{A-dagger}) translates into
\begin{eqnarray}               \label{P-dagger}
  P_g^\dagger = R_g P_{g^{-1}}  \qquad  (\forall g \in G) \; .
\end{eqnarray}
Using the $\star$-involution instead, we obtain
\begin{eqnarray}
  P_g^\dagger = R_{g^{-1}} P_g   \qquad  (\forall g \in G) \; .
\end{eqnarray}

\vskip.3cm
\noindent
{\em Remark.} Gauge theory on finite groups has been discussed in
previous work by Sitarz \cite{Sita92}. His results are not quite
in accordance with ours. Moreover, our approach stresses the fact
that a group structure is just an auxiliary structure which can be
used to deal with the differential calculus and gauge theory on $G$
in a more convenient way.

\section{The case $G=\Ir_N$}
\setcounter{equation}{0}
In this section we study the differential calculus on a finite set $G$
of $N$ elements with the help of the group structure of $\Ir_N$.
This leads to another look at the matrix representation of the
differential calculus in section 5.1. The example of gauge theory on
$\Ir_2$ is discussed in section 5.2 making contact with Connes'
two-point space model \cite{Conn}.
\vskip.3cm

If we describe $\Ir_N$ as the set of numbers $\lbrace 0,1,\ldots, N-1
\rbrace$ with addition modulo $N$ as the group structure, then
the functions $x_m, m=0,\ldots,N-1$ defined by
\begin{eqnarray}
                x_m (n) = \delta_{m,n}
\end{eqnarray}
correspond to the functions $x_g$ of section 2.
Let us introduce a new function
\begin{eqnarray}         \label{y}
     y := \sum_{n=0}^{N-1} \, q^n \, x_n
\end{eqnarray}
with $q \in \Cx$ a primitive $N$-th root of unity, i.e. $q^N =1$. Then
\begin{eqnarray}                 \label{y^n}
     y^n = \sum_{m=0}^{N-1} \, q^{mn} \, x_m
\end{eqnarray}
and, in particular, $y^N = \idty$. Note that the last equation
describes the $N$-point set in the simplest possible algebraic way.
It replaces the set of equations (\ref{sum=1}) and (\ref{xx-x}).
\vskip.2cm

Like the $x_n, n=0, \ldots,N-1,$
the set of functions $y^0, \ldots y^{N-1}$ also span the algebra
$\cal A$ of functions on $\Ir_N$. Using the identity
\begin{eqnarray}
    \sum_{m=0}^{N-1} \, q^{nm}
     = N \, \delta_{n,0}
\end{eqnarray}
(\ref{y^n}) can be inverted,
\begin{eqnarray}
     x_n = {1 \over N} \, \sum_{m=0}^{N-1} \, q^{-mn} \, y^m \; .
\end{eqnarray}
A function $f$ on $G$ can be written as
\begin{eqnarray}       \label{f=sum-y}
                f = \sum_{n=0}^{N-1} \, y^n \, f_n
\end{eqnarray}
where
\begin{eqnarray}
     f_n = {1 \over N} \, \sum_{m=0}^{N-1} \, q^{-mn} \, f(m) \; .
\end{eqnarray}
\vskip.2cm

The two involutions introduced in section 2 act on $y$ as follows,
\begin{eqnarray}           \label{y-invol}
     y^\ast = y^{-1} \quad , \quad  y^\star = y    \; .
\end{eqnarray}
\vskip.2cm

We shall now rewrite the differential calculus introduced in section
2 in terms of $y^n$. (\ref{f-dx}) implies
\begin{eqnarray}             \label{y-dy^n}
 y \, dy^n + dy \, y^n &=& dy^{n+1}   \qquad (n=1, \ldots, N-2) \\
 y \, dy^{N-1} + dy \, y^{N-1} &=& 0   \; .
\end{eqnarray}
The 1-forms $\theta_g$ introduced in section 4 now take
the form
\begin{eqnarray}
     \theta_n = {1 \over N} \, \sum_{m=1}^{N-1} \, q^{-mn} \,
                    dy^m \, y^{-m}  \; .
\end{eqnarray}
The right action on $\cal A$ is given by $(R_n f)(y) = f(q^n y)$.
(\ref{f-theta}) then implies\footnote{Acting with $d$
on a relation like (\ref{y-theta_n}) one should expect to obtain
additional relations (commutation relations for 1-forms). This is not
so in our case. The relations (\ref{y-theta_n}) are simply
consequences of the general setting
of differential calculus as given by (\ref{didty})-(\ref{Leibniz}).}
\begin{eqnarray}                 \label{y-theta_n}
    y \, \theta_n = q^n \, \theta_n \, y
        \qquad (n=1, \ldots, N-1)
\end{eqnarray}
i.e., for each $n \neq 0$ we have the algebra of a `quantum plane'
\cite{Mani88,Wess+Zumi90}. The well-known finite-dimensional
representations of the quantum plane (for $q$ a root of unity) lead
us again to matrix representations of the differential calculus
(see section 5.1).
\vskip.2cm

Using (\ref{f=sum-y}), the differential of a function $f$ is
given by
\begin{eqnarray}
     df =  \sum_{m=0}^{N-1} \, d(y^m) \, f_m      \; .
\end{eqnarray}
It may appear strange that the differentials $dy, dy^2, \ldots$ are
linearly independent although $y^k$ depends algebraically on $y$.
Indeed, we shall see in section 6.1 that an additional condition
can be imposed on the universal differential calculus which
`corrects' this affair.

\subsection{On matrix representations of the differential calculus}
The finite-dimensional representations of the `quantum plane' algebra
subject to the relation
\begin{eqnarray}
            a \, b = q \, b \, a
\end{eqnarray}
(where $q$ is an $N$-th primitive root of unity)
are given up to equivalence by the $N\times N$-matrices
\begin{eqnarray}
a = \left(\begin{array}{cccc}
   1      & 0      &\cdots & 0 \\
   0      & q      &       & 0 \\
   \vdots &        &\ddots &\vdots \\
   0      & 0      &\cdots & q^{N-1} \end{array}\right)
                                                           \quad
b = \left(\begin{array}{cccccc}
         0      & 0      & 0     & \cdots & 0      & 1      \\
         1      & 0      & 0     & \cdots & 0      & 0      \\
         0      & 1      & 0     & \cdots & 0      & 0      \\
         \vdots & \vdots &\ddots &\ddots  & \vdots & \vdots \\
         0      & 0      &\cdots &  1     & 0      & 0      \\
         0      & 0      &\cdots &  0     & 1      & 0
          \end{array}\right)
\end{eqnarray}
which are known to generate the whole algebra $M_N(\Cx)$ of
complex $N\times N$-matrices \cite{Morr67} (see also \cite{q-reps}).
They satisfy
\begin{eqnarray}
                   a^N = {\bf 1} = b^N
\end{eqnarray}
where ${\bf 1}$ denotes the $N\times N$ unit matrix. In terms of
$a$ and $b$ we can now represent the differential calculus by
\begin{eqnarray}
   \rho(y) = \left(\begin{array}{cc}
             a & 0 \\  0 & a
             \end{array}\right)
\end{eqnarray}
and
\begin{eqnarray}         \label{theta-rep}
  \rho(\theta_k) = \left(\begin{array}{cc}
   0 & \sum_{\ell = 0}^{N-1} c_{k \ell} \, a^\ell \; b^k \\
   \sum_{\ell = 0}^{N-1} c'_{k \ell} \, a^\ell \; b^k & 0
                   \end{array}\right)
\end{eqnarray}
where $c_{k \ell}$ and $c'_{k \ell}$ are complex numbers.
\vskip.3cm

The $\ast$-involution acts on matrices by
Hermitian conjugation. The equation (\ref{theta-ast}), i.e.
$\theta_k^\ast = - \theta_{N-k}$, then leads to the condition
\begin{eqnarray}                 \label{c-ast}
    c^\ast_{k \ell} = - q^{k \ell} \, c'_{N-k, N-\ell}
\end{eqnarray}
for the constants in (\ref{theta-rep}).
\vskip.3cm

The $\star$-involution acts on a matrix $B$ such that
$B^\star = P \, B^\ast \, P$ where the matrix $P$ has entries
$P_{ij} = \delta_{i, N-j}$. Then $a^\star = a$ as required by
(\ref{y-invol}) and also $b^\star = b$. From $\theta_k^\star =
-\theta_k$ we now get the condition
\begin{eqnarray}            \label{c-star}
    c^\ast_{k \ell} = - q^{k \ell} \, c'_{k \ell}
\end{eqnarray}
(where $\ast$ denotes complex conjugation).
\vskip.3cm

More general matrix representations of the differential calculus are
given by
\begin{eqnarray}
  \rho(y) = \left(\begin{array}{cc}
            {\bf 1}_M \otimes a & 0 \\
                              0 & {\bf 1}_M \otimes a
                  \end{array}\right)
    \; , \quad
  \rho(\theta_k) = \left(\begin{array}{cc}
   0 & \sum_{\ell =0}^{N-1} c_{k \ell} \otimes a^\ell \, b^k \\
   \sum_{\ell =0}^{N-1} c'_{k \ell} \otimes a^\ell \, b^k & 0
         \end{array}\right)  	\quad
\end{eqnarray}
where the $c_{k \ell}$ and $c'_{k \ell}$ are now $M \times
M$-matrices. In this case $\ast$ in (\ref{c-ast}) and (\ref{c-star})
has to be understood as Hermitian conjugation.

\subsection{Gauge theory on $\Ir_2$}
The simplest nontrivial example of a discrete space is a two-point
space, of course. This can be endowed with the group structure
of $\Ir_2$. In this case we have $q=-1$, $y^2 = \idty$ and
\begin{eqnarray}
                y \, d y = - dy \, y   \; .
\end{eqnarray}
The two involutions introduced in section 2 coincide in the case
under consideration.
The field strength of an anti-Hermitian connection $A$ takes the form
\begin{eqnarray}
 F  = \theta_1 \theta_1 \, \lbrack (R_1 P_1) \, P_1 -1
    \rbrack
    = (\theta_1)^2 \, \lbrack P_1^\dagger P_1 -1
    \rbrack
\end{eqnarray}
where (\ref{P_e=-1}) and (\ref{P-dagger}) have been used.
The two-form $(\theta_1)^2 = - {1 \over 4} (dy)^2$
commutes with all $f \in {\cal A}$. As a consequence, the
transformation law $F'=U F U^{-1}$ is shared by the coefficient
function of $F$. We can therefore build a gauge-invariant Lagrangian,
\begin{eqnarray}
 {\cal L} := \mbox{Tr} (F^\dagger F)
           = (\theta_1)^4 \, \mbox{Tr} \lbrack P_1^\dagger P_1 -1
             \rbrack^2
\end{eqnarray}
where Tr denotes the ordinary matrix trace. In order to construct
an action, we need a kind of integral, a trace tr acting on forms.
Such an integral should only have nonzero values on forms which
commute with all functions $f \in \cal A$. Using the properties
which tr has in a representation (see appendix A) one
finds
\begin{eqnarray}        \label{Z2-action}
 S := \mbox{tr} {\cal L}
    = \mbox{tr} (\theta_1^4) \; 2 \,
     \mbox{Tr} \lbrack \Phi^\dagger \Phi - 1 \rbrack^2
\end{eqnarray}
where we have set $\Phi:= P_1(0)$ and used $P_1(1) = \Phi^\dagger$
which follows from (\ref{P-dagger}). The constant $\mbox{tr}
(\theta_1^4)$  plays the role of a coupling constant.
(\ref{Z2-action}) shows that the Yang-Mills action on a two-point
space is nothing but the usual Higgs potential, a crucial observation
made in \cite{Conn}. It becomes a field on a manifold $\cal M$ when
the formalism is extended to ${\cal M} \times \Ir_2$ (see also
\cite{Sita93}).

\section{Reductions of the universal differential calculus}
\setcounter{equation}{0}
So far we have dealt with the `universal' differential calculus
on $\cal A$, i.e. we have only used the general rules
(\ref{didty})-(\ref{Leibniz}) of differential calculus.
There is some freedom to impose additional conditions which are
consistent with the universal differential calculus.
\vskip.2cm

When the differential calculus is formulated in terms of the
cyclic function $y$ introduced in section 5,
there is a natural choice for such a condition in the form of a
commutation relation between $y$ and its differential.
This is elaborated in section 6.1 and generalized in section 6.2.
\vskip.2cm

Section 6.3 contains a general discussion of reductions of the
universal differential calculus which in particular shows
that additional relations can be imposed on the differential
calculus without changing its dimensionality (i.e., the number
of linearly independent differentials).

\subsection{From the universal differential calculus to a calculus on a
lattice}
The relation
\begin{eqnarray}            \label{y-q-plane}
                y \, dy = q \, dy \, y   \; .
\end{eqnarray}
leads to a consistent differential calculus on the algebra generated
by $y$ \cite{Mani91}. For $N>2$ this condition is not consistent
with the $\ast$-involution (for which $y^\ast = y^{-1}$). It is
consistent, however, if we choose the $\star$-involution for
which $y^\star = y$. From (\ref{y-q-plane}) we deduce
\begin{eqnarray}
             d y^n = \sum_{m=0}^{n-1} \, y^m \, dy \, y^{n-m-1}
                   = \lbrack n \rbrack_q \, dy \, y^{n-1}
\end{eqnarray}
where
\begin{eqnarray}
               \lbrack n \rbrack_q := {1 - q^n \over 1 - q}     \; .
\end{eqnarray}
In particular,
\begin{eqnarray}
                d y^N = \lbrack N \rbrack_q \, dy \, y^{N-1} = 0
\end{eqnarray}
(since $q^N = 1$) in accordance with $y^N = \idty$.
This suffices to conclude that (\ref{y-q-plane}) gives a consistent
reduction of our universal differential calculus.
(\ref{y-dy^n}) is indeed identically satisfied.
\vskip.2cm

The differential of a function $f$ is now given by
\begin{eqnarray}        \label{df-qderiv}
          d f =  dy \, { f(q y) - f(y) \over (q-1) \, y }
\end{eqnarray}
which involves the so-called $q$-derivative, and the 1-form
$\theta_0$ takes the form\footnote{If we consider the differential
calculus with (\ref{y-q-plane}) on the algebra of functions on $\Rl$
where $q$ is not a root of unity, the 1-form on the right hand side
plays a special role as a measure for integrating functions of $y$.
The corresponding integral sums the values of the function
on a $q$-lattice (cf equation $(4.20)$ in \cite{DMH92}).}
\begin{eqnarray}
          \theta_0 =  {1 \over 1-q} \, dy \, y^{-1}     \; .
\end{eqnarray}
Applying $d$ to (\ref{y-q-plane}) leads to $(d y)^2 = 0 $.
\vskip.2cm
\noindent
{\em Remark.} For the ordinary differential calculus where
$y \, dy =  dy \, y$ we have $dy^N = N \, y^{N-1} dy$ which is
{\em not} consistent with $y^N = \idty$. Commutative algebras are
therefore not in general compatible with the ordinary differential
calculus.
\vskip.3cm
\noindent
In \cite{DMH92,DMHS93} we have considered a certain
deformation of the ordinary differential calculus on a manifold.
In one dimension, the deformation can be expressed in the form
\begin{eqnarray}             \label{latt-calc}
                 \lbrack X , dX \rbrack = dX \, a
\end{eqnarray}
where $X$ is a coordinate function on $\Rl$ and $a$ is a positive
real constant. An action for a (classical) field theory can be
formulated in terms of the deformed differential calculus and turns
out to describe a corresponding lattice theory where $a$ plays the
role of the lattice spacing. In terms of the new coordinate
$y = q^{x/a}$ with $q \in \Cx$, $q \neq 1$, the commutation relation
(\ref{latt-calc}) is transformed into (\ref{y-q-plane}) (see
\cite{DMH92} for details).
If $q$ is an $N$-th root of unity, we are considering a closed
(periodic) lattice of $N$ points instead of a lattice on the real line.
\vskip.2cm
\noindent
A differential calculus with $M$ linearly independent differentials on
a set $G$ of order $N$ should be thought of as associating $M$
dimensions with it.
In the case of the universal differential calculus, the differential
$df$ of a function $f$ on $G$ involves -- as `partial derivatives' --
the differences of the values of $f$ at pairs of points according to
(\ref{df=sum-difference}).\footnote{(\ref{df=sum-difference})
actually only involves the differences $f(g)-f(e)$ which suggests to
only draw lines from $e$ to the other points of the set. Note, however,
that the choice of $e$ is arbitrary.}
In this sense this differential calculus gives the structure of an
($N-1$)-dimensional polyhedron in $N$ dimensions to the set $G$ (where
the $N$ points of $G$ appear as the vertices).
\vskip.2cm

One also arrives at such a picture in the following way. Let $x_n$
be coordinate functions on $\Rl^N$. We may then consider the
equations (\ref{sum=1}) and (\ref{xx-x}) as algebraic equations imposed
on the functions $x_n$. Their solutions determine a set of $N$
points in $\Rl^N$ which form the vertices of an ($N-1$)-dimensional
polyhedron (see Fig.1).

\unitlength1.cm
\begin{picture}(9.,7.)(-6.,-2.5)
\put(0.1,0.1) {0}
\thicklines
\put(0.,0.) {\vector(1,0){4.}}
\put(0.,0.) {\vector(0,1){4.}}
\put(0.,0.) {\vector(-1,-1){2.3}}
\put(0.,2.) {\circle*{0.3}}
\put(2.,0.) {\circle*{0.3}}
\put(-1.2,-1.2) {\circle*{0.3}}
\thinlines
\put(-1.2,-1.2){\line(2,5){1.3}}
\put(-1.2,-1.2){\line(5,2){3.}}
\put(0.,2.){\line(1,-1){2.}}
\end{picture}

\centerline{
\begin{minipage}[t]{9cm}
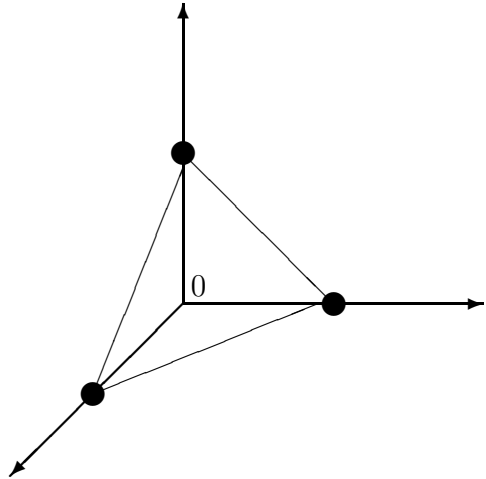
\centerline{\bf Fig.1}
\noindent
\small
The geometric structure induced on the $N=3$ point set by the universal
differential calculus.
\end{minipage}    }

\vskip.5cm
\noindent
It is hard to see how one should formulate and understand the
reduction of the differential calculus in terms of the `coordinates'
$x_n$ (cf section 6.3, however). The reformulation in terms of the
single function $y$ makes
it easy to formulate a constraint, i.e. (\ref{y-q-plane}), which reduces
the $N-1$ dimensions of the universal differential calculus to a single
one. The corresponding geometric picture (based on (\ref{df-qderiv}))
is obtained by drawing the set of $N$-th roots of unity in the
(complex) plane (see Fig.2).

\unitlength1.cm
\begin{picture}(5.,5.)(-7.4,-2.5)
\thicklines
\put(-2.,0.) {\vector(1,0){4.}}
\put(0.,-2.) {\vector(0,1){4.}}
\put(0.,1.) {\circle*{0.2}}
\put(1.,0.) {\circle*{0.2}}
\put(-1.,0.) {\circle*{0.2}}
\put(0.,-1.) {\circle*{0.2}}
\put(0.71,0.71) {\circle*{0.2}}
\put(-0.71,0.71) {\circle*{0.2}}
\put(-0.71,-0.71) {\circle*{0.2}}
\put(0.71,-0.71) {\circle*{0.2}}
\thinlines
\put(0.,1.){\line(5,-2){0.6}}
\put(0.,1.){\line(-5,-2){0.6}}
\put(0.,-1.){\line(5,2){0.6}}
\put(0.,-1.){\line(-5,2){0.6}}
\put(-1.,0.){\line(1,2){0.4}}
\put(-1.,0.){\line(1,-2){0.4}}
\put(1.,0.){\line(-1,2){0.4}}
\put(1.,0.){\line(-1,-2){0.4}}
\end{picture}

\centerline{
\begin{minipage}[t]{9cm}
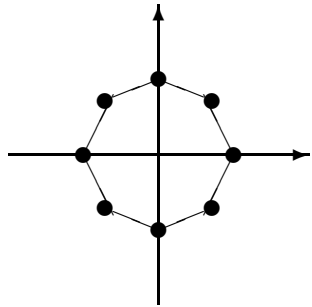
\centerline{\bf Fig.2}
\noindent
\small
The $N=8$ point set as a one-dimensional closed lattice embedded in
the two-dimensional plane (the structure given to it by the
one-dimensional differential calculus).
\end{minipage}    }

\subsection{$G = \Ir_{N_1} \times \cdots \times \Ir_{N_r}$ reductions}
In section 5 we considered the group structure $\Ir_N$ on a set
of $N$ elements. The formulae given there are easily generalized,
using multi-index notation, to a group structure $G = \Ir_{N_1}
\times \cdots \times \Ir_{N_r}$ where $N = N_1 \cdots N_r$.
Let $\underline{n} \in G$ and
\begin{eqnarray}
      y^{\underline{n}} &:=& \prod_{k=1}^r y_k^{n_k}
      \quad , \quad        y_k^{N_k} = \idty         \\
      q^{\underline{n}} &:=& \prod_{k=1}^r q_k^{n_k}
      \quad , \quad        q_k^{N_k} = 1
\end{eqnarray}
($q_k$ primitive roots). Then
\begin{eqnarray}
  y^{\underline{m}} \, dy^{\underline{n}} + dy^{\underline{m}} \,
  y^{\underline{n}} = dy^{\underline{m}+\underline{n}}    \; .
                    \label{y-dy-multi}
\end{eqnarray}
Generalizing the differential calculus reduction scheme of section
6.1, we impose the commutation relations
\begin{eqnarray}          \label{y-crs-multi}
   y_k \, dy_\ell = q_k^{\delta_{k \ell}} \, dy_\ell \, y_k  \; .
\end{eqnarray}
As a consequence, we have
\begin{eqnarray}
   dy^{\underline{n}} = \sum_{k=1}^r \, \lbrack n_k \rbrack_{q_k}
     \, dy_k \, y^{\underline{n}-\underline{e}_k}
\end{eqnarray}
where $\underline{e}_k \in G$ has components $e_{k \ell} =
\delta_{k \ell}$. Furthermore,
\begin{eqnarray}
  y^{\underline{m}} \, dy^{\underline{n}} = \sum_{k=1}^r \,
  \lbrack n_k \rbrack_{q_k} \, q_k^{m_k} \, dy_k \,
  y^{\underline{m}+\underline{n}-\underline{e}_k}     \;.
\end{eqnarray}
(\ref{y-crs-multi}) has to be compatible with (\ref{y-dy-multi}).
Using the last two equations in (\ref{y-dy-multi}), we obtain
\begin{eqnarray}
  0 = \sum_{k=1}^r \, \left( \lbrack n_k \rbrack_{q_k} \, q^{m_k}
      + \lbrack m_k \rbrack_{q_k} - \lbrack m_k + n_k \rbrack_{q_k}
      \right) \, dy_k \, y^{\underline{m}+\underline{n}
      -\underline{e}_k}
\end{eqnarray}
which is identically satisfied since the expression in round
brackets on the right hand side vanishes identically. This
shows that (\ref{y-crs-multi}) indeed defines a consistent reduction
of the universal differential calculus.
\vskip.3cm
\noindent
The geometric structure which the reduced differential calculus
places on the $N$-point set is a `discrete torus'. It is a
cartesian product of `discrete circles' like the one shown in Fig.2.
\vskip.3cm
\noindent
{\em Example:} $\Ir_2 \times \Ir_3$   \\
The constraints
\begin{eqnarray}
                u^2= \idty \quad , \quad v^3 = \idty
\end{eqnarray}
imposed on the two coordinates $u,v$ on $\Rl^2$ determine a set of
six points. In terms of $u$ and $v$, the universal differential calculus
on the six-point space is given by the following set of rules,
\begin{eqnarray}
\begin{minipage}[c]{6cm}
\begin{eqnarray*}
      u \, du + du \, u & = & 0                   \\
      u \, dv + du \, v & = & d(uv)               \\
      u \, d(uv) + du \, u v & = & dv             \\
      u \, d(v^2) + du \, v^2 & = & d(uv^2)       \\
      u \, d(uv^2) + du \, uv^2 & = & d(v^2)
\end{eqnarray*}
\end{minipage}
\begin{minipage}[c]{6cm}
\begin{eqnarray*}
      v \, du + dv \, u & = & d(uv)               \\
      v \, dv + dv \, v & = & d(v^2)              \\
      v \, d(uv) + dv \, u v & = & d(uv^2)        \\
      v \, d(v^2) + dv \, v^2 & = & 0             \\
      v \, d(uv^2) + dv \, u v^2 & = & du    \; .
\end{eqnarray*}
\end{minipage}
\end{eqnarray}    \\
The 1-form $\theta_0$ takes the form
\begin{eqnarray}
 \theta_0 = {1\over6} \, \lbrack du \, u + dv \, v^2 + d(uv) \, u v^2
            + d(v^2) \, v + d(uv^2) \, u v \rbrack   \; .
\end{eqnarray}
Any function $f$ on the set can be written as
\begin{eqnarray}
     f = \sum_{i=0}^1 \sum_{j=0}^2 u^i \, v^j \, f_{ij}
\end{eqnarray}
with constants $f_{ij}$. This leads to the expression
\begin{eqnarray}             \label{df-uv}
     df = \sum_{i=0}^1 \sum_{j=0}^2 d(u^i v^j) \, f_{ij}  \; .
\end{eqnarray}
for its differential.
The reduction is now performed by imposing the relations
\begin{eqnarray}
  v \, dv = p \, dv \, v    \quad , \quad
  v \, du = du \, v         \quad , \quad
  u \, dv = dv \, u         \label{uv-rels}
\end{eqnarray}
where $p$ is a cubic primitive root of unity. They imply
\begin{eqnarray}
    d(uv)  & = & du \, v + dv \, u                   \\
    d(v^2) & = & (1+p) \, dv \, v      \label{dv^2}  \\
    d(uv^2) & = & du \, v^2 + (1+p) \, dv \, u \, v    \; .
\end{eqnarray}
Using these relations in (\ref{df-uv}) one finds
\begin{eqnarray}         \label{df-uv-red}
               df = du \, {f(u,v)-f(-u,v) \over 2u}
                    + dv \, {f(u,pv)-f(u,v) \over (p-1) v}   \; .
\end{eqnarray}
Furthermore,
\begin{eqnarray}
 \theta_0 = {1\over2} \, du \, u^{-1} - {1\over p-1} \, dv \, v^{-1}
\end{eqnarray}
and we obtain the 2-form relations
\begin{eqnarray}
     du \, dv + dv \, du = 0  \quad , \quad  dv \, dv = 0  \; .
\end{eqnarray}
In the sense of our discussion in section 6.1, via (\ref{df-uv-red})
the reduced differential calculus gives a two-dimensional torus
structure to the set of six points.

\subsection{Further remarks about reductions}
So far we have understood a {\em reduction} of a differential calculus
as a procedure to reduce its dimensionality (i.e., the number of
linearly independent differentials). This is done by adding extra
relations to the differential calculus. It is, however, possible
to add relations without changing the dimension.
\vskip.2cm

In section 2 we have introduced the 1-forms $\theta_{ij}$ and
shown that setting one (or several) of these forms to zero does not
lead to any further constraint on the first order differential
calculus. It leads, of course, to 2-form relations $d \theta_{ij}=0$
and additional conditions according to the involution which is
used.
\vskip.2cm

With a differential calculus we can associate a directed graph
with $N$ vertices. An arrow connects vertex $i$ with vertex $j$
whenever $\theta_{ij} \neq 0$. For the universal differential
calculus this means that each pair of vertices is connected by
two lines with opposite direction. We can represent this graph
by the $N\times N$-matrix which has zeros on the diagonal and
all other entries equal to 1. A one in the $i$-th row and
$j$-th column stands for an arrow pointing from vertex $i$ to
vertex $j$.
Setting $\theta_{12} =0$, for
example, means that we have to delete the arrow from vertex 1
to vertex 2 (in the corresponding matrix, the 1 in the first
row and second column has to be replaced by 0).
Imposing also $\theta_{21}=0$ separates the two
vertices (which may, however, still be connected via other
vertices).
\vskip.3cm
\noindent
{\em Example.} In section 6.2 a reduction of the universal
differential calculus on the six-point space to a
two-dimensional differential calculus was considered.
Using
\begin{eqnarray}
   u &=& x_0 - x_1 + x_2 -x_3 + x_4 - x_5   \\
   v &=& x_0 + p^2 \, x_1 + p \, x_2 + x_3 + p^2 \, x_4 + p \, x_5
\end{eqnarray}
(where $p$ is a primitive cubic root of unity), the reduction
conditions (\ref{uv-rels}) can be expressed in the form $\theta_{ij}=0$
for certain values of the indices $i,j$. These equations are
summarized in the matrix
\begin{eqnarray}
         \left(\begin{array}{cccccc}
         0 & 0 & 0 & 1 & 1 & 0   \\
         0 & 0 & 0 & 0 & 1 & 1   \\
         1 & 0 & 0 & 0 & 0 & 1   \\
         1 & 1 & 0 & 0 & 0 & 0   \\
         0 & 1 & 1 & 0 & 0 & 0   \\
         0 & 0 & 1 & 1 & 0 & 0
          \end{array}\right)             \; .
\end{eqnarray}
The corresponding graph is shown in Fig.3.
That each row and each column of the matrix have precisely two 1's
is related to the fact that the reduced differential
calculus is two-dimensional. For the graph it means that each
vertex has two incoming and two outgoing arrows.
In general, the reduction procedure leads to graphs where the number
of incoming and outgoing arrows varies from vertex to vertex. This
somehow means that the dimension varies over the set of points.

\unitlength2.cm
\begin{picture}(5.,2.5)(-3.6,-0.8)
\thicklines
\put(0.,1.) {\circle*{0.1}}
\put(1.,0.) {\circle*{0.1}}
\put(-1.,0.) {\circle*{0.1}}
\put(0.,1.) {\vector(-1,-1){1.}}
\put(-1.,0.) {\vector(1,0){2.}}
\put(1.,0.) {\vector(-1,1){1.}}
\put(0.,-.3) {\circle*{0.1}}
\put(1.,.7) {\circle*{0.1}}
\put(-1.,.7) {\circle*{0.1}}
\put(0.,-.3) {\vector(1,1){1.}}
\put(1.,.7) {\vector(-1,0){2.}}
\put(-1.,.7) {\vector(1,-1){1.}}
\put(-.7,.6) {\vector(3,-1){1.5}}
\put(-.7,.6) {\vector(-3,1){.1}}
\put(-.7,.1) {\vector(3,1){1.5}}
\put(-.7,.1) {\vector(-3,-1){.1}}
\put(0.,-.1) {\vector(0,1){1}}
\put(0,-.1) {\vector(0,-1){.1}}
\put(-1.2,0.) {4}
\put(-1.2,.7) {5}
\put(0.,1.2) {0}
\put(1.2,.7) {1}
\put(1.2,0.) {2}
\put(0.,-.6) {3}
\end{picture}

\centerline{
\begin{minipage}[t]{9cm}
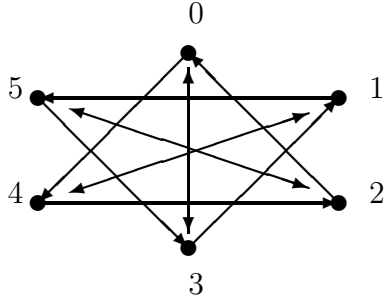
\centerline{\bf Fig.3}
\noindent
\small
The graph associated with a two-dimensional reduction of the
universal differential calculus on a six-point space.
\end{minipage}    }

\section{Conclusions}
\setcounter{equation}{0}
We have formulated differential calculus and gauge theory on
an arbitrary finite set. Endowing the latter with a group structure,
one can define analogues of Maurer-Cartan forms (see also
\cite{Sita92}). It it then
straightforward to define principal fiber bundles with discrete
structure groups and connections on them. Other differential
geometric structures are expected to also have a `discrete analogue'.
\vskip.2cm

In the special example of gauge theory on a two-point space we
recovered the geometric interpretation of the Higgs field as in Connes'
formulation of the standard model (\cite{Conn}, see also \cite{Sita93}).
\vskip.2cm

The universal differential calculus (in the sense of Connes)
associates with all linearly independent elements of an algebra
corresponding linearly independent differentials.
This means that it assigns a geometric picture (a polyhedron) in $N$
dimensions to a set of $N$ elements.
The universal differential calculus admits reductions to consistent
differential calculi with which one can associate a similar geometric
picture in lower dimensions. We have only given examples which
certainly do not exhaust the possibilities.
In particular, we discussed a reduction to a single dimension in
section 6.1. In \cite{DMH92} the resulting differential calculus has
been shown to be equivalent to a `noncommutative' differential calculus
on an equidistant periodic lattice.
\vskip.2cm

There is one point which we would like to stress here. In an algebraic
sense, one can easily construct reductions of the universal
differential calculus. The problem is that, in general, one is not
able to find some geometric picture associated with such a reduction.
Such a picture may be found by expressing the reduced calculus in
suitable `coordinates'. However, we do not have a systematic way yet
to find such coordinates. On the other hand, we have shown
that associated with certain group structures there are choices of
coordinates in terms of which reductions of the universal
differential calculus lead to a geometric understanding of the
resulting differential calculus.
\vskip.2cm

It is important, however, to keep the following in mind. When
we speak about a `geometric picture' we are actually guided by
continuum geometry, thinking of spheres, tori etc.. A finite
set of points can be connected in such a way that the resulting
structure reminds us of such a continuum geometric picture.
But there are other connection structures for which no
corresponding continuum picture exists. This means that there are
many more possibilities for discrete structures. This also
concerns the dimensionality. To a set of points we can assign
different dimensions. The fact that for a discrete set there is
no rigid notion of dimension has some interesting aspects.
If a space-time model is set up in such a framework, the
dimension may even change with length scale and in such a way
incorporate features of Kaluza-Klein theories (cf \cite{BLMS87}).
\vskip.2cm

In \cite{DMH92,DMHS93} we were interested in deformations of the
ordinary differential calculus on the algebra of functions on $\Rl^N$.
In case of a certain deformation (cf (\ref{latt-calc})) it turned out
that the differential calculus could be restricted to functions on a
lattice. In the same way we can understand each of the differential
calculi of the present paper as a calculus on $\Rl^M$ when the
calculus has $M$ independent differentials. The calculus still
contains the information about the point set in the
following way. Consider, for example, the differential calculus with
$u \, du = - du \, u$ where $u$ is a real coordinate on $\Rl$. Using
the Leibniz rule, it implies $d(u^2)=0$ which means that $u^2$ is
a constant with respect to the differential calculus under
consideration. It follows that the `constants' are precisely the
even functions $h$ of $u$ (i.e. $h(-u)=h(u)$). Since every function
$f$ can be written in a unique way as $f=h_0 + h_1 u$ with even
functions $h_0,h_1$, it is represented by a pair of `constants'
and in this sense we have a two-point space.
\vskip.2cm

As already mentioned in the introduction, there are several
approaches towards physical theories based on discrete spaces in
the literature. Noncommutative differential geometry of discrete
spaces should have some impact on these approaches and vice versa.

\vskip.5cm
\noindent
{\bf Acknowledgment.}  A. D. is grateful to the Heraeus-Foundation
for financial support during a visit at the Institute for Theoretical
Physics in G\"ottingen. We are also grateful to John Madore for
pointing out an error in the first version of this paper.

\begin{appendix}
\renewcommand{\theequation} {\Alph{section}.\arabic{equation}}

\section{A representation of the differential calculus on $\Ir_2$}
\setcounter{equation}{0}
Following Connes' treatment of the two-point space \cite{Conn},
we represent a function $f$ on $\Ir_2$ by a diagonal matrix
\begin{eqnarray}
      f = \left( \begin{array}{cc} f(0) \, {\bf 1} & 0 \\
                                   0 & f(1) \, {\bf 1}
                 \end{array} \right)
\end{eqnarray}
where $\bf 1$ denotes the $m \times m$ unit matrix. The differential
of $f$ is represented by
\begin{eqnarray}
     d f = \lbrack f(0) - f(1) \rbrack \; i \, \left(
           \begin{array}{cc} 0 & -M^\dagger \\
                             M &   0        \end{array} \right)
\end{eqnarray}
with a complex $m \times m$ matrix $M$. In particular, the function
$y$ introduced in section 5 is given by
\begin{eqnarray}
   y = \left( \begin{array}{cc} {\bf 1} & 0 \\
                                0 & - {\bf 1}
                 \end{array} \right)
\end{eqnarray}
and we find
\begin{eqnarray}
   d y = 2 \, i \, \left( \begin{array}{cc} 0 & - M^\dagger \\
                                   M & 0
                 \end{array} \right)
   \quad , \quad   dy^2 = 0    \; .
\end{eqnarray}
For the 1-form $\theta_1$ we obtain
\begin{eqnarray}
   \theta_1 = - {1 \over 2} \, dy \, y
                = - i \, \left( \begin{array}{cc}
                                0 &  M^\dagger \\
                                M &     0
                                \end{array} \right)
\end{eqnarray}
so that
\begin{eqnarray}
  (\theta_1)^4 =  \left( \begin{array}{cc}
                                (M^\dagger M)^2 & 0 \\
                                0 & (M M^\dagger)^2
                                \end{array} \right)
\end{eqnarray}
and therefore
\begin{eqnarray}
    \mbox{tr} \lbrack (\theta_1)^4 \, f \rbrack
 =  2 \, \mbox{tr} (M^\dagger M)^2  \; \lbrack f(0) + f(1) \rbrack
\end{eqnarray}
where tr is the ordinary matrix trace.

\section{Matrix algebras as differential algebras}
\setcounter{equation}{0}
Let $M_N$ denote the algebra of complex $N\times N$ matrices and
$\Omega$ the direct sum $M_N \oplus M_N$. The latter becomes an
algebra with the multiplication rule $(A,B)(A',B')=(AA',BB')$.
There is a natural $\Ir_2$ grading.
We call an element of $\Omega$ `even' if it is of the form $(A,A)$
and `odd' if it has the form $(A,-A)$. Since
$(A,B) = ((A+B)/2 , (A+B)/2) + ((A-B)/2 , -(A-B)/2)$, $\Omega$ splits
into a direct sum, $\Omega=\Omega^+\oplus\Omega^-$.
Let ${\cal A}_N$ denote the commutative subalgebra of $\Omega^+$
consisting of elements of the form $f=(F,F)$ where $F \in M_N$ is
diagonal. In terms of the matrices $E_{ij}$ with components
$(E_{ij})_{k\ell}=\delta_{ik} \delta_{j\ell}$ we have
\begin{eqnarray}
     f=\sum_{i=0}^{N-1} f_i \, (E_{ii},E_{ii}) \; .
\end{eqnarray}
For $i\neq j$ we introduce $\theta_{ij}:=c_{ij} \, (E_{ij},-E_{ij})$
where $c_{ij}\in \Cx$. Furthermore, we define
\begin{eqnarray}
     df := [ f , \vartheta ] \quad , \quad
   \vartheta := \sum_{i\neq j} \theta_{ij}
\end{eqnarray}
and $d(f_0 df_1\cdots df_r) := df_0 \, df_1 \cdots df_r$. With these
definitions, $\Omega$ has the structure of a differential calclulus
over ${\cal A}_N$.
\vskip.2cm

For $x_i:=(E_{ii},E_{ii})$ we find $\sum_{i=0}^{N-1} dx_i =0$ and
$ dx_i \, x_j = \theta_{ij}$ for $i\neq j$ (cf section 2).
There are $N-1$ independent differentials $dx_i,\; i=1,\ldots,N-1$
in this calculus. Setting some of the $c_{ij}$ to zero, it is
possible to reduce the number of independent differentials. In this
way one can turn $M_N$ itself into a differential algebra.
As an example, let us consider $M_3$. Let ${\cal A}_3$ be the
subalgebra of diagonal matrices and
\begin{eqnarray}
   \vartheta := \left(\begin{array}{ccc} 0 & 0     & c_{02} \\
                                         0 & 0     & c_{12} \\
                                    c_{20} & c_{21}& 0
                      \end{array} \right)
\end{eqnarray}
(i.e., we set $c_{01}=c_{10}=0$). With the above definition of
$d$, $M_3$ becomes a differential algebra $\Omega_3$ over
${\cal A}_3$. In this case, we have a $\Ir_2$ grading
$\Omega_3 = \Omega^+ \oplus \Omega^-$ where $\Omega^+$ and
$\Omega^-$ consist, respectively, of the matrices of the form
\begin{eqnarray}
   \left(\begin{array}{ccc} \ast & \ast & 0 \\
                            \ast & \ast & 0 \\
                            0    & 0    & \ast  \end{array}\right)
   \quad \mbox{and} \quad
   \left(\begin{array}{ccc} 0    & 0    & \ast \\
                            0    & 0    & \ast \\
                            \ast & \ast & 0  \end{array}\right)
\end{eqnarray}
with possible non-zero entries indicated by a $\ast$.

\end{appendix}

\end{document}